\documentclass[prl,aps,twocolumn]{revtex4} 
\usepackage{epsfig} 
\usepackage{bm} 
\newcommand{\B}[1]{{\bm{#1}}} 
\newcommand{\C}[1]{{\mathcal{#1}}} 
\renewcommand{\it}[1]{\textit{#1}} 
\newcommand{\Onecol} {\begin{widetext} \onecolumngrid} %% 2 -> 1 
\newcommand{\Twocol} {\end{widetext} \twocolumngrid} %% 1 -> 2 
 
\newcommand{\be}{\begin{equation}} 
\newcommand{\ba}{\begin{array}} 
\newcommand{\bea}{\begin{eqnarray}} 
\newcommand{\bfi}{\begin{figure}} 
\newcommand{\ee}{\end{equation}} 
\newcommand{\ea}{\end{array}} 
\newcommand{\eea}{\end{eqnarray}} 
\newcommand{\efi}{\end{figure}}

\newcommand{\ra}{\right\rangle} 
\newcommand{\la}{\left\langle} 
\begin{document} 
\title{A simple model for drag reduction} 
\author{Roberto Benzi$^{1}$ and Itamar Procaccia$^{2,3}$} 
\affiliation{$^1$ Dip. di Fisica and INFM, Universit\`a ``Tor 
Vergata", Via della Ricerca Scientifica 1, I-00133 Roma, Italy.\\
$^2$ Dept. of Chemical Physics, The Weizmann Institute of Science, Rehovot, 
76100 Israel. \\$^3$ Centro Intranacional de Ciencias, UNAM, Cuernavaca, Mexico} 
%\pacs{47.27-i, 47.27.Nz, 47.27.Ak} 
\begin{abstract} 
Direct Numerical Simulations established that the FENE-P model of 
viscoelastic flows 
exhibits the phenomenon of turbulent drag reduction which is caused in 
experiments 
by dilute polymeric additives. To gain 
analytic understanding of the phenomenon we introduce in this Letter a 
simple 1-dimensional 
model of the FENE-P equations. We demonstrate drag reduction in the 
simple model, and explain analytically the main observations which include 
(i) reduction 
of velocity gradients for fixed throughput and (ii) increase of throughput 
for fixed dissipation. 
\vskip 0.2cm 
\end{abstract} 
\maketitle 
%%%%%%%%%%%%%%%%%%%%%%%%%%%%%%%%%%%%%%%%%%%%%%%%%%%%%%%%%%%%%%%%%%%%%%%%%%%% 
%% 
The addition of few tens of parts per million (by weight) of long-chain 
polymers 
to turbulent fluids can bring 
about a reduction of the friction drag by up to 80\% \cite{00SW}. This 
``drag reduction" 
phenomenon has important practical implications besides being interesting 
from the 
fundamental point of view, integrating turbulence research with polymer 
physics. 
In spite of intense interest for an extended period of time 
\cite{69Lu,75Virk,90Ge}, 
Sreenivasan and White \cite{00SW} recently concluded that ``it is fair to 
say that the extensive - 
and continuing - activity has not produced a firm grasp of the mechanisms of 
drag reduction". 
Recently however it was shown that drag reduction is observed in Direct 
Numerical 
Simulation of model viscoelastic hydrodynamic equations 
\cite{97THKN,98DSB,00ACP}. From the 
theoretical viewpoint these observations are crucial, indicating that the 
phenomenon is included in the solutions of the model equations. 
Understanding drag reduction 
then becomes a usual challenge of theoretical physics. In this Letter we 
present 
a further simplification of the model equations and gain analytic insights 
into the phenomenon. 
The FENE-P equation for the fluid velocity $\B u(\B r,t)$ contains an 
additional stress tensor related to the polymer: 
\begin{equation} 
\frac{\partial \B u}{\partial t}+(\B u\cdot \B \nabla) \B u=-\B \nabla p 
+\nu_s \nabla^2 \B u +\B \nabla \cdot \B {\C T}+\B F\ , \label{Equ} 
\end{equation} 
where $\nu_s$ is the viscosity of the neat fluid, $\B F$ is the forcing and 
the stress tensor 
$\B {\C T}$ is determined by the polymer conformation tensor $\B R$ 
according to 
\begin{equation} 
\B {\C T}(\B r,t) = \frac{\nu_p}{\tau_p}\left[\frac{f(\B r,t)}{\rho_0^2} \B 
R(\B r,t) -\B 1 \right] \ . 
\end{equation} 
here $\nu_p$ is a viscosity parameter, $\tau_p$ is a relaxation time for the 
polymer 
conformation tensor, $\rho_0$ is the rms extension of the polymers in 
equilibrium, 
and $f(\B r,t)$ is a function that limits the growth of the trace of 
$\B R$. The model is closed by the equation of motion for the conformation 
tensor which reads 
\begin{eqnarray} 
\frac{\partial R_{\alpha\beta}}{\partial t}+(\B u\cdot \B \nabla) 
R_{\alpha\beta} 
&&=\frac{\partial u_\alpha}{\partial r_\gamma}R_{\gamma\beta} 
+R_{\alpha\gamma}\frac{\partial u_\gamma}{\partial r_\beta}\label{EqR}\\ 
&&-\frac{1}{\tau_p}\left[ f(\B r,t) R_{\alpha\beta} -\rho_0^2 
\delta_{\alpha\beta} \right]\nonumber 
\end{eqnarray} 
These equations were simulated on the computer in a channel or pipe 
geometry. The main observations 
on the effect of the polymer on the turbulent flow that we need to focus on 
are the following: 
%%%%%%%%%%%%%%%%%%%%%%%%%%% 
\begin{figure} 
\centering 
\includegraphics[width=.45\textwidth]{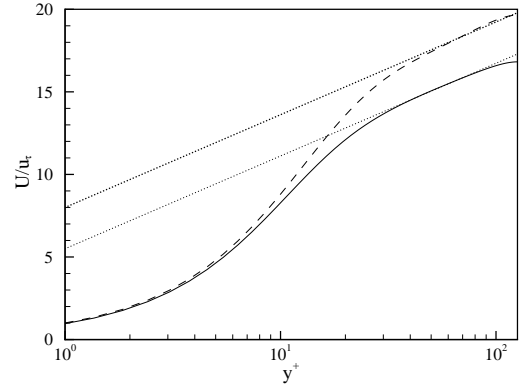} 
\caption{The mean flow velocity as a function of the 
distance from the 
wall for the FENE-P (dashed line) vs. the Newtonian flow (continuous line). The profiles
hardly  change near 
the wall, but the amplitude is larger for the FENE-P solution.} 
\label{fig1} 
\end{figure} 
%%%%%%%%%%%%%%%%%%% 
(i) For a fixed pressure gradient at the wall the fluid throughput is 
increased (see 
Fig. \ref{fig1}). 
(ii) For a fixed throughput the gradient at the wall decreases (i.e. the 
dissipation decreases). 
(iii) The trace of the conformation tensor $\B R$ follows qualitatively the 
rms streamwise velocity 
(see Fig. 2). We are particularly interested in point (iii) since in our 
opinion the space dependence of 
the amount of stretching (and with it of the effective viscosity) is 
crucial, and compare 
\cite{01GLP,02GLP} for a discussion of this point in the context of the 
instability of laminar flows. 
%%%%%%%%%%%%%%%%%%%%%%%%%%% 
\begin{figure} 
\centering 
\includegraphics[width=.35\textwidth]{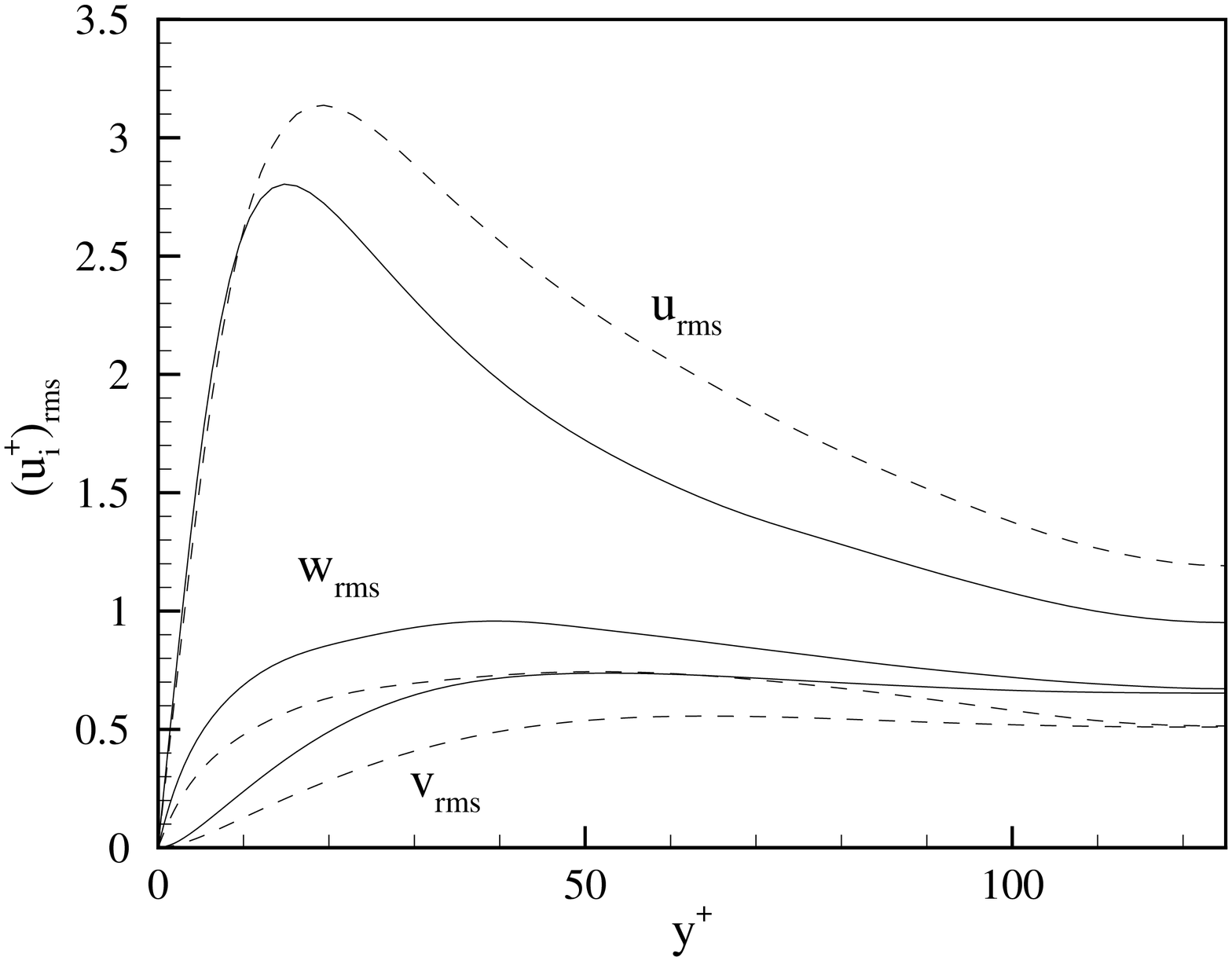} 
\includegraphics[width=.35\textwidth]{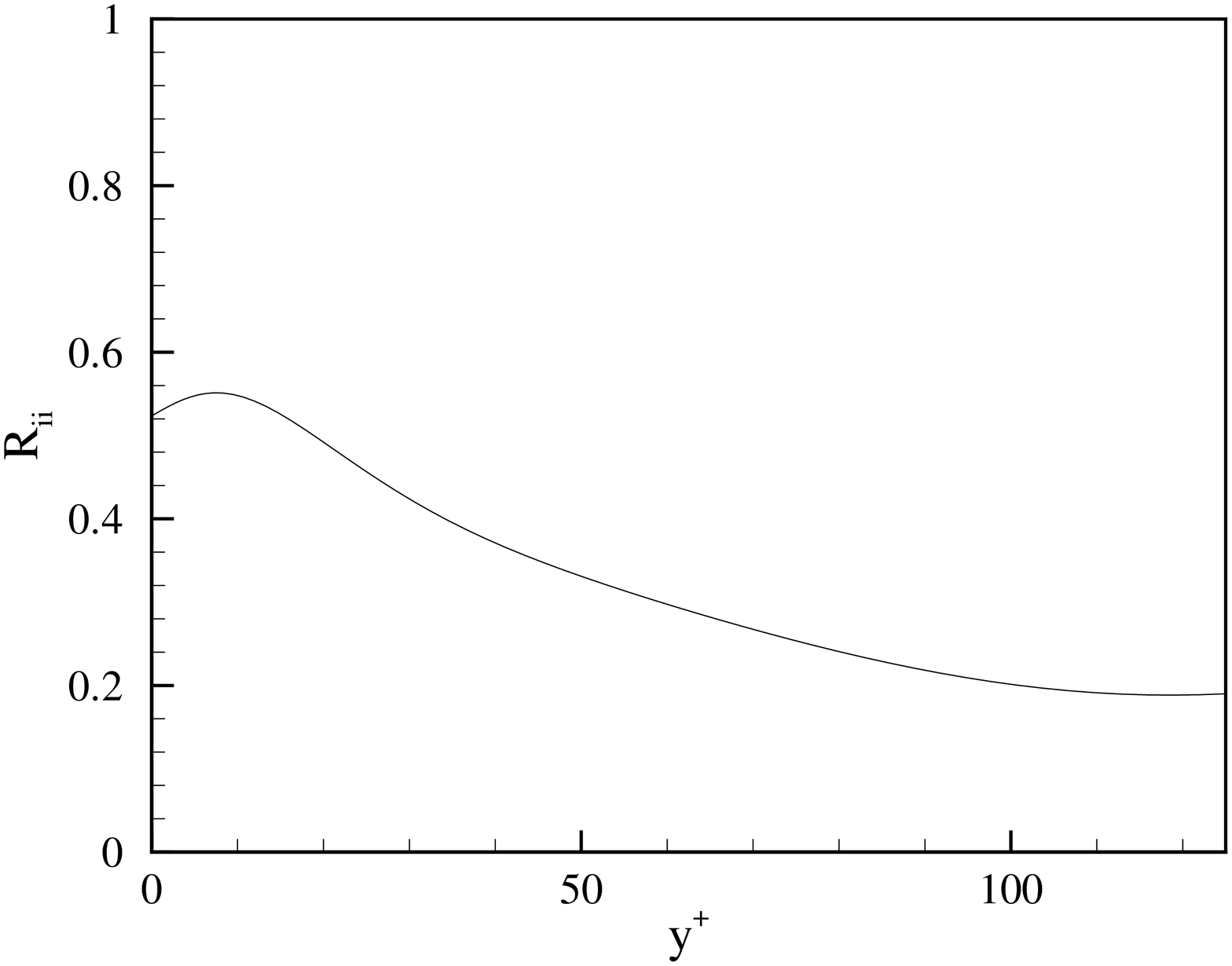} 
\caption{Upper panel: the dependence of the rms velocity fluctuations as a 
function 
of the distance from the wall. We are interested in $U_{\rm rms}$ for 
comparison with 
our model. Lower panel: the trace of the conformation tensor $\B R$ as a 
function 
of the distance from the wall. We stress the qualitative similarity to the 
dependence 
of $U_{\rm rms}$ in the upper panel.} 
\label{fig2} 
\end{figure} 
%%%%%%%%%%%%%%%%%%% 
Obviously, Eqs. (\ref{Equ})-(\ref{EqR}) as they stand are not amenable to 
analytic investigation 
in the turbulent regime. To gain insight we therefore attempt to simplify 
them as much as 
possible without losing the main phenomena (i)-(iii). Consider therefore a 
model for the 
streamwise velocity which is the Burger's equation ($u$ in the streamwise 
directions 
with gradients in the $y$ (wall-normal) direction), to which the effect of a 
scalar $R$ 
is added: 
\begin{eqnarray} 
u_t + u u_y &=& \nu u_{yy} + s R_y + F \ , \label{u}\\ 
R_t + u R_y &=& -\frac{1}{\tau} R + R u_y \ . \label{R} 
\end{eqnarray} 
where a subscript $y$ stands for a partial derivative with respect to $y$. 
In the following we shall denote 
Eqs. (\ref{u})-(\ref{R}) with the acronimous {\it {uR model}}. 
The parameter $s$ is related to the polymer concentration, and 
$\tau$ is the relaxation time of $R$. We will consider the model in the 
domain 
$-L\le y\le L$, with boundary conditions chosen later. We will denote 
spatial averages by pointed brackets, $\langle A\rangle\equiv \int_{-L}^L 
A(y) dy$. 
The simplicity of the {\it uR} model allows us to state the energy budget in 
simple terms. Multiplying (\ref{u}) by $u$ and taking 
the spatial average of (\ref{u}) and (\ref{R}) we obtain: 
\be 
\frac{1}{2}\frac{d}{dt} \la u^2\ra = - \nu \la u_y^2 \ra + s \la u R_y \ra 
+ \la F u \ra 
\label{eu} 
\ee 
\be 
\frac{d}{dt} \la R\ra = -\frac{1}{\tau} \la R \ra + 2\la R u_y \ra 
\label{eR} 
\ee 
The term $ \langle u R_y\rangle $ measures the ``energy" given by the 
velocity field $u$ to the polymer field 
$R$. Multiplying (\ref{eR}) by $s/2$ and summing (\ref{eR}) with (\ref{eu}) 
we obtain: 
\be 
\frac{d}{dt} ( \frac{1}{2}\la u^2 \ra + \frac{s}{2} \la R \ra ) 
= - \nu \la u_y^2 \ra - \frac{s}{2\tau}\la R\ra + \la Fu\ra 
\label{energy} 
\ee 
In the steady state the overall power $\la Fu\ra$ is balanced by the 
overall energy dissipation per unit time $D$, 
$ D = \nu \la u_y^2 \ra + \frac{s}{2\tau}\la R\ra$. 
The term $ \frac{1}{2}\la u^2 \ra + \frac{s}{2} \la R \ra $ 
represents the sum of the kinetic energy of the flow 
plus the potential energy of the stretched polymers. 
We remark that already from these elementary consideration it becomes clear 
that for a fixed power input the existence of the term $\frac{s}{2\tau}\la 
R\ra$ 
necessarily reduces the gradients of $u$ in agreement with point (ii) above. 
To address points (ii) and (iii) further we consider next the solution of 
the model with 
$F=0$ and with a fixed velocity $u$ and stretching $R$ at $-L$ and $L$. In 
other words, we take as boundary 
conditions 
\begin{equation} 
u(-L) = u_0,\, u(L) = - u_0,\, R(-L)= R(L) = R_0 \ . \label{case1} 
\end{equation} 
In Fig. 3 we compare the solution of the {\it uR} model to that of the pure 
Burger's equation (i.e. 
Eq. (\ref{u}) with $R=F=0$.) To focus our thinking we would like the reader 
to 
consider the solution in the left half space as a model of the streamwise 
velocity 
component in the lower half channel, with the solution in the right half 
space being 
simply an anti-symmetric copy. The position of the lateral ``wall" is 
modeled by 
the point where $u=0$. 
Thinking this way points (ii) and (iii) are clearly 
demonstrated. We proceed now analytically to demonstrate drag reduction 
(point (ii)) and to understand 
the profile of $R$ (point (iii)). 
%%%%%%%%%%%%%%%%%%%%%%%%%%%% 
\begin{figure} 
\centering 
\includegraphics[width=.35\textwidth]{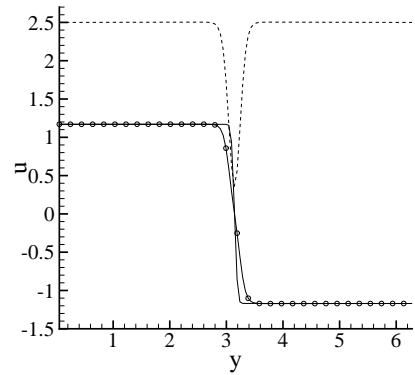} 
\caption{Comparison of the solution of the {\it uR} model (connected circles) to 
the solution of Burger's 
equation (continuous line). The dashed line corresponds to the 
function $R$. 
The parameters are $\tau = 10^4$, $u_0 = 1.17$, $s = 0.25$ and 
$R_0 = 2.5$.} 
\label{fig3} 
\end{figure} 
%%%%%%%%%%%%%%%%%%%%% 
First we consider the stationary solution of the pure Burger's equation. 
Integrate equation (\ref{u}) in $y$ 
to find 
\be 
\frac{1}{2} u^2 = \nu u_y + \frac{1}{2}u_0^2 
\label{c} 
\ee 
where the constant of integration was fixed by noticing that for $L$ 
sufficiently large $u_y$ is expected 
to vanish at the boundaries. Multiplying (\ref{c}) by $u_y$, and integrating 
between $-L$ 
and $L$ using the boundary conditions, we find the viscous dissipation 
$\epsilon$ 
\begin{equation} 
\epsilon\equiv \nu\langle u_x^2\rangle = \frac{2}{3} u_0^3 
\label{dissu} 
\end{equation} 
Next we consider the solution of the {\it uR} model for the same boundary 
conditions (\ref{case1}) and $R \ge 0$. 
In the stationary state $R_t = 0 $, and by dividing Eq. (\ref{R}) by $Ru$ we 
can integrate it formally in $y$ and 
obtain: 
\be 
R = a |u| exp(-\frac{1}{\tau}\int \frac{dy}{u}) 
\label{Rtau} 
\ee 
where $a$ is a constant of integration. This equation is the explanation of 
point (iii). 
It says that for small velocity $ u \sim 0$, i.e. for $y \sim 0$, 
$R$ {\em necessarily goes to $0$}. In particular, approximating $u = -my$ 
near the point $y=0$, we obtain $R \sim |u|^b$ 
where 
$b = 1 +1/(m\tau)$. Thus we should expect that at positions with small $u$ 
where the gradient 
of $u$ is large the generic behavior of $R$ is a cusp with $R=0$ for $m\tau 
\ge 1$. 
To compute the dissipation analytically we consider the limit $\tau 
\rightarrow \infty$, 
i.e. we look for a solution at the zero order of the perturbation series in 
$1/\tau$. In this limit: 
\be 
R = \frac{R_0}{u_0} |u|. 
\label{Ru} 
\ee 
Returning to Eq. (\ref{u}) we integrate it in $y$ to obtain 
\begin{equation} 
\nu u_y =\frac{1}{2} u^2 -SR +sR_0 - \frac{1}{2} u_0^2 \ . 
\end{equation} 
We can now substitute Eq. (\ref{Ru}) in the domain $-L \le y\le 0$ where 
$|u|=u$, 
and integrate between $-L$ and $0$. Multiplying the result by a factor of 2 
we find 
the viscous dissipation $\epsilon_R$ 
\begin{equation} 
\epsilon_R= \frac{2}{3} u_0^3 -3 s R_0 u_0 
\end{equation} 
This result is an analytic demonstration of point (ii). 
We note that 
our analysis has been performed in the limit $\tau \rightarrow \infty$. For 
large but finite values of $\tau$ the 
qualitative picture we have drawn is unchanged. 
Needless to say, the above discussion can be reformulated by keeping 
constant 
the energy dissipation while increasing the value of $u$ at the boundary, to 
demonstrate 
point (i). We choose however to demonstrate point (i) next, using a forced 
solution. 
Point (i) is most clearly demonstrated in the {\it uR} model 
using periodic boundary conditions and constant forcing. 
We consider $0\le y \le 2\pi$ and choose the external forcing $F$ to be: 
\begin{eqnarray} 
F(y) &=& f_0 sin(4y/3) \ , \quad \text{for}~ 0\le y \le \frac{3\pi}{2} \ , 
\nonumber\\ 
F(y) &=& f_1 sin(4y) \ , \quad \quad \text{for}~ \frac{3\pi}{2}\le y \le 
2\pi \ . \label{forcing} 
\end{eqnarray} 
We examined the solutions of the {\it uR} model for the set of parameters 
$f_0 = 0.1$, $f_1 = 0.05$, $\nu = 0.01$, $ s = 
0.01$ and initial conditions 
$R(y) = 2 sin(y)$. The remaining parameter is the relaxation time $\tau$. It 
turns out that 
for very small values of $\tau$, no effect of the polymer field is observed. 
(We remark that for periodic boundary conditions the limit $\tau \rightarrow 
0$ corresponds to the 
case of no polymer.) For 
$\tau \rightarrow \infty$ no stationary solutions can be obtained. For 
$\tau$ smaller 
than some critical value $ \tau_c$, the solution 
of the {\it uR} model shows stable stationary solutions with drag reduction. 
The typical situation is presented in 
Fig. \ref{fig4}, showing the numerical solutions for $\tau= (0.15)^{-1} \le 
\tau_c$, compared against the solution of 
Burger's equation. The {\it uR} model shows a larger amplitude near the 
strongest shock due to forcing at $x = 3\pi/4$. 
It is worth noting that the gradient is maintained extremely close to the 
one obtained by the Burger equation, 
demonstrating nicely point (i). 
%%%%%%%%%%%%%%%%%%%%%%%%%%% 
\begin{figure} 
\centering 
\includegraphics[width=.35\textwidth]{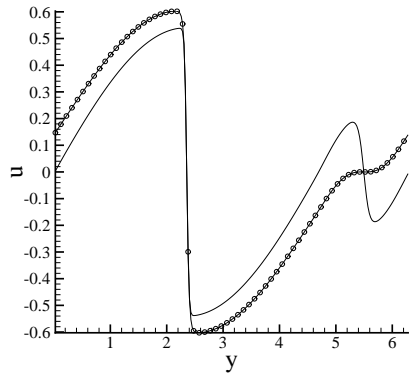} 
\caption{The solution of the {\it uR} model with constant 
forcing and periodic boundary conditions (connceted circles). 
 Continuous line: the Burger 
equation without the polymer.} 
\label{fig4} 
\end{figure} 
%%%%%%%%%%%%%%%%%%% 
Point (iii) is nicely demonstrated in 
Fig. \ref{fig5} which presents the solution for $R$ together with $u^2$ for 
both the {\it uR} model and the Burger 
equation. As one can clearly see, the behavior of $R$ is similar to what 
observed in Fig. \ref{fig1}, namely there is 
a qualitative similarity between the space dependence of $R$ and $u^2$, here 
with sharp cusp in $R$ near the point of 
maximum gradient of $u$. On the other hand, the smallest shock present in 
the solution of Burger's equation has been 
completely smoothed out by the {\it uR} model. This is an indication that 
when $R$ is not sufficiently suppressed 
where the gradient of $u$ is significant, there can be {\em drag 
enhancement}. This important point will be 
addressed again in the concluding remarks. 
%%%%%%%%%%%%%%%%%%%%%%%%% 
\begin{figure} 
\centering 
\includegraphics[width=.35\textwidth]{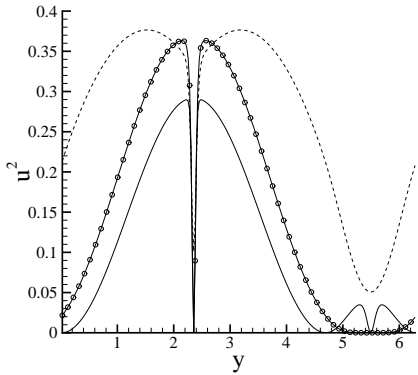} 
\caption{The solution of the {\it uR} model with constant 
forcing and periodic boundary conditions. Dashed line: 
$R(y)$. Connected circles: $20u^2$ of the solution 
of {\it uR}. Solid line: $20 u^2$ of the solution of Burger's equation.} 
\label{fig5} 
\end{figure} 
%%%%%%%%%%%%%%%%%%% 
Again, the simplicity of the model affords an analytic explanation of 
why the solution 
near the biggest shock shows a larger velocity amplitude compared to 
the Burger equation. Let $y_0$ be the position of the maximal velocity 
near the shock. The position $y_0$ is unchanged in the two models. 
We can expand $u$, $F$ and $R$ as power series near $y_0$. 
Let $\Delta$ be defined as $\Delta = y_s -y_0$, where $y_s = 
3\pi/4$ is where the velocity vanishes (the ``wall" position in our model). 
We have: 
\be 
u(y) = u_0 - u_2 y^2 \ , F(y) = F_0 + F_1 y \ , R(y) = R_0 + R_1 y 
\label{expansion} 
\ee 
where $F_0 \sim f_0 \Delta$ and $F_1 \sim -f_0 $ because of Eq. 
(\ref{forcing}). 
Inserting (\ref{expansion}) into the {\it uR} model, we obtain: 
\be 
u_0 = \frac{\nu f_0 + sR_0\tau^{-1}}{\Delta f_0} = u_b (1+P) , \;\;\; P = 
\frac{sR_0}{\nu \tau f_0} 
\label{solution} 
\ee 
where $u_b = \nu/\Delta$ is the solution obtained without polymer, ($P=0$). 
The increment of the velocity, which is responsible for the drag reduction, 
is proportional to $sR_0$. 

In summary, we have introduced a simple model of the effect of polymeric 
additives to 
Newtonian fluids, with the aim of understanding in simple mathematical terms 
some 
of the prominent features associated with the phenomenon of drag reduction. 
Needless 
to say, the model cannot be taken as quantitative; we were concerned with 
the 
qualitative features summarized above for convenience as point (i)-(iii). We 
demonstrated 
that our model reproduces these qualitative features, and provided 
straightforward analytic 
explanation to all those features. It appears that we can draw from the 
results of this 
model a few important conclusions: 
(i) Arguments concerning the turbulent cascade process do not appear 
essential. 
These arguments are the hallmark of the theory presented in \cite{90Ge} 
which 
proposed that the main effect of the polymer is to introduce a dissipative 
cutoff at scales larger than the Kolmogorov scale, due to 
the polymer relaxation time matching there the hydrodynamic time scale. 
Looking at Eq. (\ref{R}) one could think 
that the largest velocity gradients could be estimated as $\langle 
|u_y|\rangle \sim 1/\tau$. 
But since $R$ can go to zero where the gradient is largest, such estimates 
cannot be made. Moreover, 
just an increase in the dissipative scale cannot account for drag reduction; 
a {\em homogeneous} increase in the effective viscosity 
should lead by itself to drag enhancement rather than reduction. (ii) Drag
reduction is a phenomenon that appears on the scales of the system size,
involving energy containing modes rather than dissipative, small scale modes \cite{02ACLPP}.
(iii) The main point appears to be the {\em space dependence} of the 
stretching  of the polymer, here modeled by the value of $R(y)$. It is crucial that $R$ 
is small 
where the velocity gradients are large. It is the space dependence of the 
effective viscosity which should be looked at as the source of drag 
reduction. A similar 
conclusion was arrived to in the context of the study of the stability of 
laminar 
flows in a channel geometry \cite{01GLP,02GLP} accept that there the space 
dependence 
of the effective viscosity had been introduced by hand. In the FENE-P 
context as 
well as in our model (and presumably in actual experiments) the space 
dependence 
appears self consistently. 
It remains to understand this self consistent build up of differential 
effective 
viscosity in the context of the much more elaborate FENE-P model. In light 
of 
the present results this appears an extremely worthwhile endeavor that will 
shed 
important light on the phenomenon of drag reduction. 
%%%%%%%%%%%%%%%%%%%%%%%%%%%%%%%%% 
\acknowledgments 
This work 
was supported in part by the European Commission 
under a TMR grant, the German Israeli Foundation, and the 
Naftali and Anna Backenroth-Bronicki Fund for Research in 
Chaos and Complexity. We thank E. de Angelis for Figs. 1 and 2 taken from her 
PhD thesis.
%%%%%%%%%%%%%%%%%%%%%%%%%%%%%%%%%% 
 

\begin{thebibliography}{99} 
\bibitem{00SW} 
K. R. Sreenivasan and C. M. White, J. Fluid Mech. {\bf 409}, 149 (2000). 
\bibitem{69Lu} 
J. L. Lumley, Ann. Rev. Fluid Mech. {\bf 1}, 367 (1969) 
\bibitem{75Virk} 
P.S. Virk, AIChE J. {\bf 21}, 625 (1975) 
\bibitem{90Ge} 
P.-G. de Gennes {\em Introduction to Polymer Dynamics}, (Cambridge, 1990). 
\bibitem{97THKN} 
J.M.J de Toonder, M.A. Hulsen, G.D.C. Kuiken and F.T.M Nieuwstadt, J. Fluid. 
Mech {\bf 337}, 193 (1997). 
\bibitem{98DSB} 
C.D. Dimitropoulos, R. Sureshdumar and A.N. Beris, J. Non-Newtonian Fluid 
Mech. {\bf 79}, 433 (1998). 
\bibitem{00ACP} 
E. de Angelis, C.M. Casciola and R. Piva, CFD Journal, {\bf 9}, 1 (2000). 
\bibitem{01GLP} 
R. Govindarajan, V.S. L'vov and I. Procaccia, Phys. Rev. Lett., {\bf 87}, 
174501 (2001). 
\bibitem{02GLP} 
R. Govindarajan, V. S. L'vov and I. Procaccia, ``Stabilization of 
Hydrodynamic Flows by 
Small Viscosity Variations", Phys. Rev. E, submitted. 
\bibitem{02ACLPP}
E. De Angelis, C.M. Casciola, V.S. L'vov, R. Piva and I. Procaccia,
``Drag reduction by polymers in turbulent channel flows: Energy redistribution between invariant
empirical modes", Phys. Rev. E, to be published.
\end{thebibliography}
\end{document}